\documentstyle[11pt,aaspp4]{article}
\slugcomment{To appear in {\it The Astrophysical Journal Letters}}

\begin{document}

\title{A Two-Dimensional, Self-Consistent Model of Galactic Cosmic Rays in the
Heliosphere}
\author{V. Florinski and J. R. Jokipii}
\affil{Department of Planetary Sciences, University of Arizona, Tucson, AZ
85721}

\begin{abstract}
We present initial results from our new two-dimensional (radius and latitude),
self-consistent model of galactic cosmic rays in the heliosphere.
We focus on the latitudinal variations in the solar wind flow caused
by the energetic particles.
\par
Among other things our results show that the cosmic rays significantly modify
the latitudinal structure of the solar wind flow downstream of the termination
shock.
Specifically, for $A>0$ (corresponding to the present solar minimum) the wind
beyond the shock is driven towards the equator, resulting in a faster wind flow
near the current sheet, while for $A<0$ the effect is reversed and the wind
turns towards the pole, with a faster flow at high latitudes.
We attribute this effect to the latitudinal gradients in the cosmic ray
pressure, caused by drifts, that squeeze the flow towards the ecliptic plane
or the pole, respectively.
\end{abstract}


\section{Introduction}
The past decade has seen a number of advances in spherically-symmetric
self-consistent models of the heliosphere that include galactic cosmic rays
(GCRs).
These have been found to have a complex effect on the heliospheric
termination shock.
First, the supersonic upstream flow is decelerated by the radial cosmic-ray
pressure gradient creating a shock precursor in addition to the
gas-pressure-mediated subshock (\cite{dru81}).
Second, the shock moves inward closer to the Sun as a result of decreased
dynamic pressure of the wind upstream (see \cite{koc88}).
Third, the flow downstream of the shock is modified as well, with the gas
pressure being higher just downstream of the shock as a result of momentum
conservation (\cite{don93}).
\par
While, originally, momentum-integrated quantities were used for the particles,
recently a more sophisticated approach has been adopted which involves solving
the full cosmic-ray transport equation (\cite{par65}).
le Roux and Ptuskin (1995) and le Roux and Fichtner (1997) published a series
of sophisticated 1-D models that included both galactic and anomalous cosmic
rays.
\par
2-dimensional simulations of the cosmic rays as {\it test particles} have been
available for 2 decades (see, e.g., \cite{jok79}; \cite{jok81}; \cite{pot85};
\cite{haa95}; \cite{jok93}).
These showed that drifts in the heliospheric magnetic field play an important
role, often more so than diffusion, in transporting the GCRs inside the
heliosphere.
In particular, for $A>0$ (northern polar magnetic field directed outward), GCRs
have enhanced access to the interior regions near the pole while at low
latitudes their propagation is governed by the outward drift within the current
sheet.
Since it's impossible to treat the drifts correctly in 1 dimension, an important
part of the physical picture is being left out of one-dimensional models.
\par
This paper presents the first results from a 2-dimensional model of the
GCR-modified heliosphere.
Unlike the recent work of Izmodenov (1997) the structure of the outer 
heliosphere is greatly simplified here (there is no external interstellar flow
and, hence, no heliopause or bow shock), however our description of cosmic-ray
transport is much more complete.
Our model should only be compared with the upwind side of the heliosphere.
The purpose of this work is to illustrate the basic physics of the effects of
the GCRs on the flow in the outer heliosphere, not to reproduce in detail
actually observed wind flow or particle spectra.
While our results are still preliminary, we believe we have discovered an
important possible effect of cosmic rays downstream of the termination shock.

\section{Cosmic-Ray-Modified Solar Wind}
Our model uses a spherical ($r-\theta$) coordinate system with symmetry about
the polar axis.
Another symmetry is about the equatorial plane, which allows us to study the
northern hemisphere only; all quantities in the southern hemisphere are their
mirror images except the magnetic field components $B_r$ and $B_{\phi}$, which
change sign across the neutral sheet.
The conservation equations for the wind, neglecting the magnetic force, are
\begin{equation}
\frac{\partial\rho}{\partial t}+\frac{\partial(\rho u_i)}{\partial x_i}=0
\end{equation}
\begin{equation}
\frac{\partial u_i}{\partial t}+u_j\frac{\partial u_i}{\partial x_j}+\frac{1}
{\rho}\frac{\partial P_g}{\partial x_i}=-\frac{1}{\rho}\frac{\partial P_c}
{\partial x_i}
\end{equation}
\begin{equation}
\frac{\partial}{\partial t}\left(\frac{\rho u^2}{2}+\rho\varepsilon\right)
+\frac{\partial}{\partial x_i}\left[\rho u_i\left(\frac{u^2}{2}+w\right)\right]
=-u_i\frac{\partial P_c}{\partial x_i},
\end{equation}
where $\rho$, $\mathbf{u}$ and $P_g$ are the wind mass density, velocity and
thermal pressure, respectively.
For a polytropic gas, internal energy $\varepsilon=P_g/\rho(\gamma-1)$ and
enthalpy $w=\gamma P_g/\rho(\gamma-1)$.
$P_c$ is the cosmic-ray pressure, determined from the distribution function $f$
discussed below.
\par
In the supersonic wind, the magnetic force is negligible.
We also find the magnetic pressure to be insignificant compared to the thermal
and the GCR pressures downstream of the shock.
However some of our simulations exhibited regions where the magnetic force could
become quite large (see Discussion).
Later revisions of our model will include the dynamic effects of the magnetic
field on the wind.
\par
Parker's (1965) transport equation is used for the cosmic-ray distribution
function $f$
\begin{equation}
\frac{\partial f}{\partial t}+u_i\frac{\partial f}{\partial x_i}+v_{d,i}
\frac{\partial f}{\partial x_i}-\frac{\partial}{\partial x_i}\left(\kappa_{ij}
\frac{\partial f}{\partial x_j}\right)=\frac{1}{3}\frac{\partial u_i}
{\partial x_i}\frac{\partial f}{\partial\ln p},
\end{equation}
where $\mathbf{v_d}$ is the particle drift velocity and $\mathbf{\kappa}$ is the
(symmetric) diffusion tensor.
For protons, the relevant drift velocity components can be written as
\begin{equation}
v_{d,r}=\frac{p c v}{3 e}\frac{1}{r\sin\theta}\frac{\partial}{\partial\theta}
\left(\sin\theta\frac{B_{\phi}}{B^2}\right)
,\;\;\;\;\;\;\;\;
v_{d,\theta}=-\frac{p c v}{3 e}\frac{1}{r}\frac{\partial}{\partial r}
\left(r\frac{B_{\phi}}{B^2}\right),
\end{equation}
where $p$ and $v$ are particle's momentum and speed, respectively and $e$ is the
elementary charge.
For diffusion we adopted a commonly-used form (e.g., \cite{jok81},
Jokipii, et al. 1993):
\begin{equation}
\kappa_{ij}=\kappa_{\perp}\delta_{ij}+\frac{\left(\kappa_{\parallel}
-\kappa_{\perp}\right)B_{i}B_{j}}{B^2}
,\;\;\;\;\;\;\;\;
\kappa_{\parallel}=\kappa_0 P^{1/2}\frac{v}{c}\frac{B_0}{|B|},
\end{equation}
where $P$ is the particle's rigidity and $B_0$, the magnetic field intensity at
1 AU distance in the ecliptic plane.
In our simulations we took $\kappa_0=1.5\cdot10^{22}\mbox{cm}^2\mbox{s}^{-1}$
and $\kappa_{\perp}=0.05\kappa_{\parallel}$.
\par
We used a uniform, latitude-independent 400 km/s solar wind as the initial
condition.
The external gas pressure was chosen such that the shock was initially a sphere
located at approximately 80 AU distance from the Sun (similar to \cite{ler97}).
Our model can readily handle non-spherical shocks.
However, this simple initial geometry helps us to isolate purely
cosmic-ray-induced behavior vs. effects caused by latitudinal dependence in the
wind dynamic pressure.
\par
The solar wind equations were solved using the second order Godunov-type
numerical scheme (\cite{col90}) and the HLLE Riemann solver of Einfeldt, et al.
(1991).
The cosmic-ray equation (4) was solved using a second order ADI split scheme
based on the scheme of McKee and Mitchell (1970).

\section{Magnetic Field and Cosmic-Ray Transport}
The current version of our model neglects the magnetic force on the wind and
Faraday's law can be solved independently.
For the azimuthal component we have
\begin{equation}
\frac{\partial B_{\phi}}{\partial t}=-\frac{1}{r}\frac{\partial\left(r u_r
B_{\phi}\right)}{\partial r}-\frac{1}{r}\frac{\partial\left(u_{\theta}B_{\phi}
\right)}{\partial\theta}.
\end{equation}
The other two components of the field are usually much smaller, except at small
radii, but they still are computed for consistency using the steady-state
version of the Faraday's law $u_r B_{\theta}=u_{\theta}B_r$.
This approach is justified because of the long evolutionary time scales of the
system.
Combined with the Maxwell's equation ${\mathbf\nabla}\cdot{\mathbf B}=0$, the
above expression can be solved for the radial field component.
\par
To account for the fluctuating component of the magnetic field observed in the
polar regions of the solar wind we included the modified latitudinal field
(\cite{jok89}, \cite{jok95}), $B^m_{\theta}$ in the overall magnetic field
structure.
Consider the motion of the magnetic field footpoints due to supergranulation or
some other random process, in the $\theta$ direction.
If we assume a simple, periodic supergranulation motion, then $B^m_{\theta}
=B^m_{\theta 0}g(r,t)$ where $g$ is a rapidly-oscillating function.
The wavelengths and periods of the fluctuations are expected to be much
smaller than the spacial and time scales of the system.
By averaging the $\theta$ component of the Faraday's law, multiplied by $g$, it
may be shown that
\begin{equation}
\frac{\partial B^m_{\theta 0}}{\partial t}=-\frac{1}{r}
\frac{\partial\left(r u_r B^m_{\theta 0}\right)}{\partial r},
\end{equation}
where we used $\langle g\rangle_{r,t}=\langle g\partial g/\partial r\rangle_{r,t}=\langle g\partial g/\partial t\rangle_{r,t}=0$
and $\langle gg\rangle_{r,t}\neq0$.
A modified field computed in this fashion enters into $B^2$ which results in
transport coefficients that are smaller by several orders of magnitude at the
pole, according to equations (5) and (6).
\par
Since the magnetic field changes polarity across the equatorial current sheet,
the drift velocity contains a delta-function.
We employed two different approaches in dealing with the current sheet.
In the first we treat the drift velocity as a boundary conditions as in
Jokipii \& Kopriva (1979).
We also carried out calculations for the wide current sheet model of
Potgieter \& Moraal (1985).
The two sets of results were consistent.
All results presented here are for a current sheet with a half-width of $8^0$.

\section{Model Results}
We have performed a number of simulations with different resolutions and
equatorial boundary conditions.
For the results presented here the spatial resolution was 1 AU in the radial
direction and $0.5^0$ in latitude.
At the outer boundary of the simulation, the gas pressure was taken to be
uniform at 1.1 $\mbox{eV cm}^{-3}$ and the GCR pressure was 0.4
$\mbox{eV cm}^{-3}$.
The cosmic-ray spectrum at the far boundary was the same as in Jokipii \&
Kopriva (1979).
The inner, cosmic-ray-absorbing, boundary was placed at 5 AU to avoid numerical
errors due to insufficient resolution in the region of the rapid wind expansion.
\begin{figure}
\epsscale{0.5}
\plotone{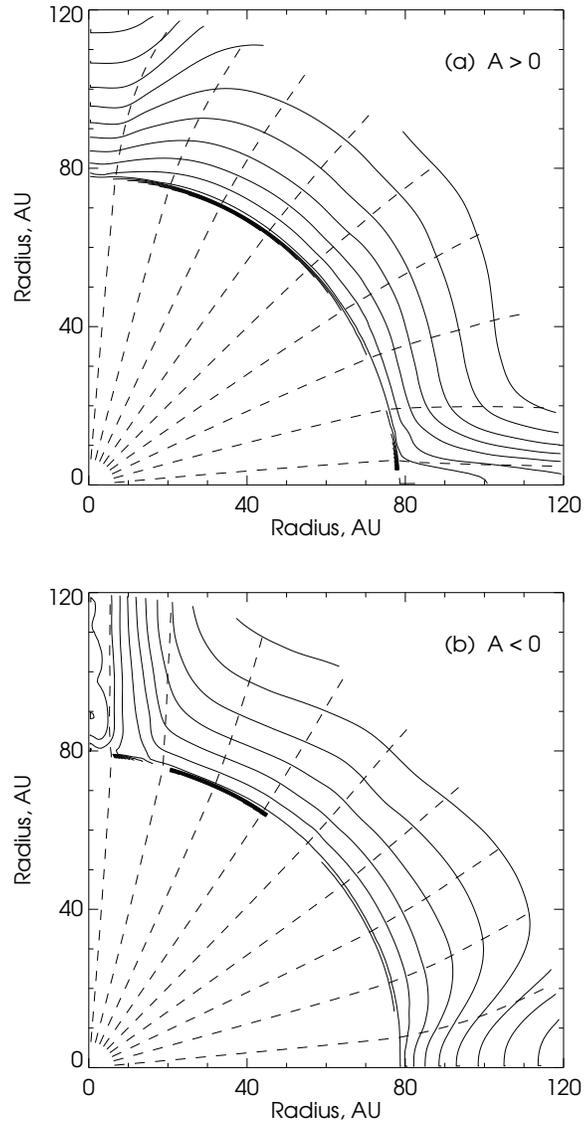}
\figcaption{(a,b) Radial solar wind velocity contours (solid lines) and
streamlines (dashed lines), for $A<0$ (a) and $A>0$ (b).
The contours shown are equally spaced between 200 km/s and 0.}
\end{figure}
Results presented here are not for a steady-state configuration.
In fact we have never achieved a real steady state except in case of the small
cosmic-ray content.
However, the rate of change of most parameters slows down after a few years.
After a sufficiently long period of time (of the order of 10 years) we begin to
see an inflow at the outer boundary at the pole for $A>0$ or the ecliptic, for
$A<0$, which cannot be handled properly by the present boundary conditions.
Because the magnetic field reversal occurs every 11 years, we present here
results for the state 8 years from the start of the simulation.
\par
Figures 1a and 1b show the radial solar-wind speed contours and streamlines for
$A>0$ and $A<0$ respectively.
We can see a significant change in the flow pattern downstream of the shock.
The wind turns away from the pole and towards the equator in the
case $A>0$ while the opposite is true for $A<0$.
Since the subsonic flow tends to be incompressible, its density varies little
at mid-latitudes, all changes are concentrated near the pole and the equatorial
neutral sheet.
Note that the shock remains essentially spherical, even though there is clearly
a strong latitudinal dependence of the various parameters.
The subshock compression ratio varies slightly between 3.8 at the equator and
4.2 at the pole (for $A>0$).
\begin{figure}
\epsscale{0.5}
\plotone{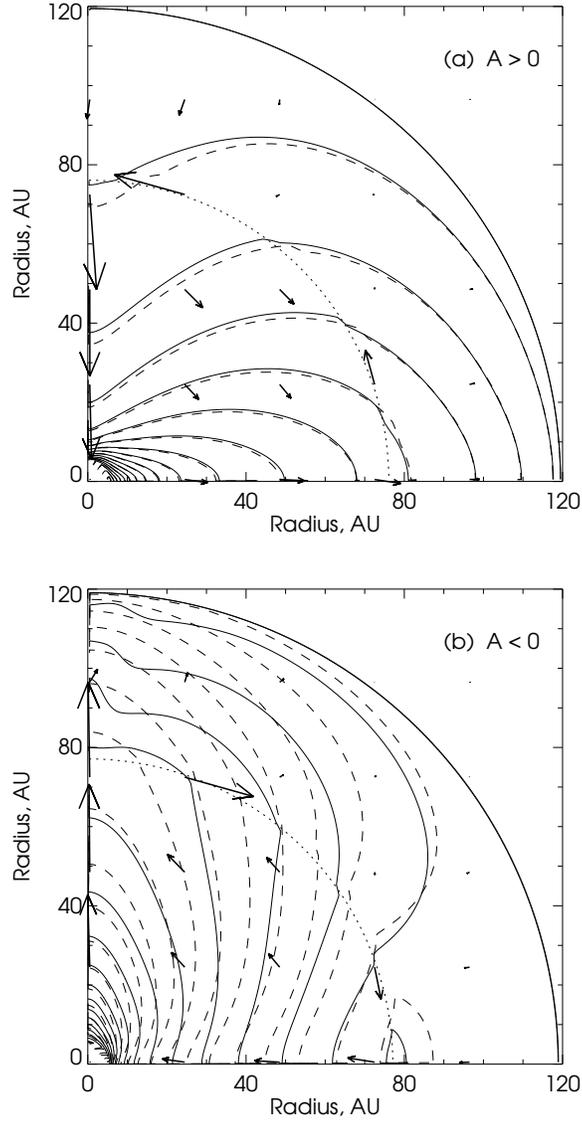}
\figcaption{(a,b) Cosmic-ray intensity contours of 1.6 GeV protons (solid
lines).
For comparison, test-particle results are plotted with dashed lines.
Case (a) is for $A>0$, case (b) is for $A<0$.
Contours are evenly spaced in 5\% increments.
Dotted lines mark the approximate location of the termination shock for the
self-consistent case.
Arrows show proton drift velocities; arrow length $\sim\sqrt{|\mathbf{v_d}|}$.}
\end{figure}
\begin{figure}
\epsscale{0.5}
\plotone{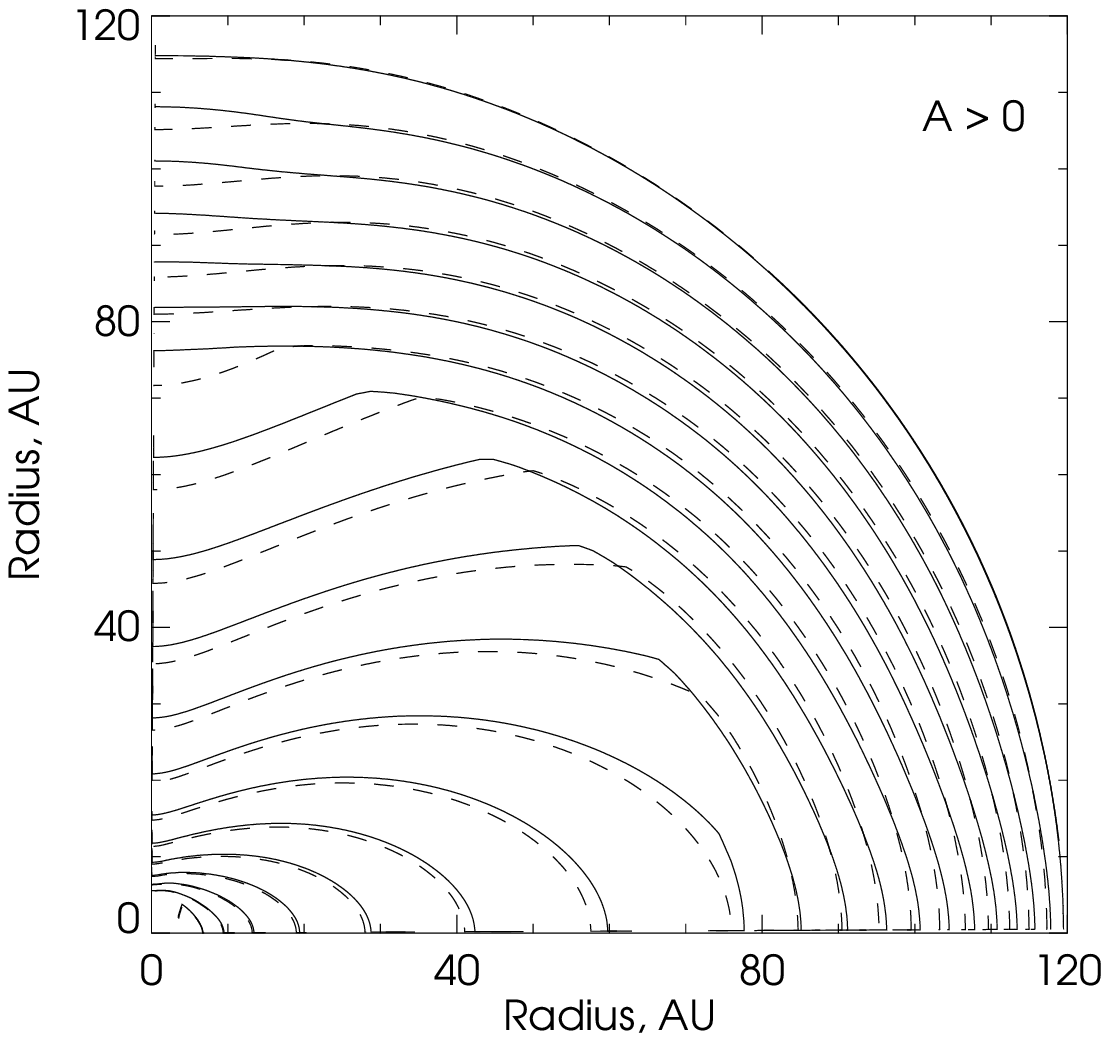}
\figcaption{Cosmic-ray intensity contours of 72 MeV protons.
Solid lines are current model results, dashed lines are test-particle results.
Case $A<0$ not shown due to space limitations.}
\end{figure}
Figures 2a and 2b show the intensity contours for 1.6 GeV protons, and Figure 3
plots the same quantity for 72 MeV protons, both figures also show
test-particle results for comparison (dashed lines).
We find that for $A>0$, the cosmic-ray distribution is only slightly affected
by the changing conditions in the solar wind.
Here radial transport coefficients become slightly smaller both near the equator
(because of compression, $B_{\phi}$ is increased) and in the polar region (due
to enhanced $B^m_{\theta}$).
The small change in intensity agrees with results of Jokipii \& Davila (1981)
who found that the particles are not strongly affected by relatively small
changes in transport coefficients.
Most of the difference we see is probably due to the changed position of the
shock.
However, for $A<0$ there is a larger difference in particle intensity at high
latitudes.
This is a result of higher degree of wind modification near the pole than in the
$A>0$ case.
\begin{figure}
\epsscale{0.6}
\plotone{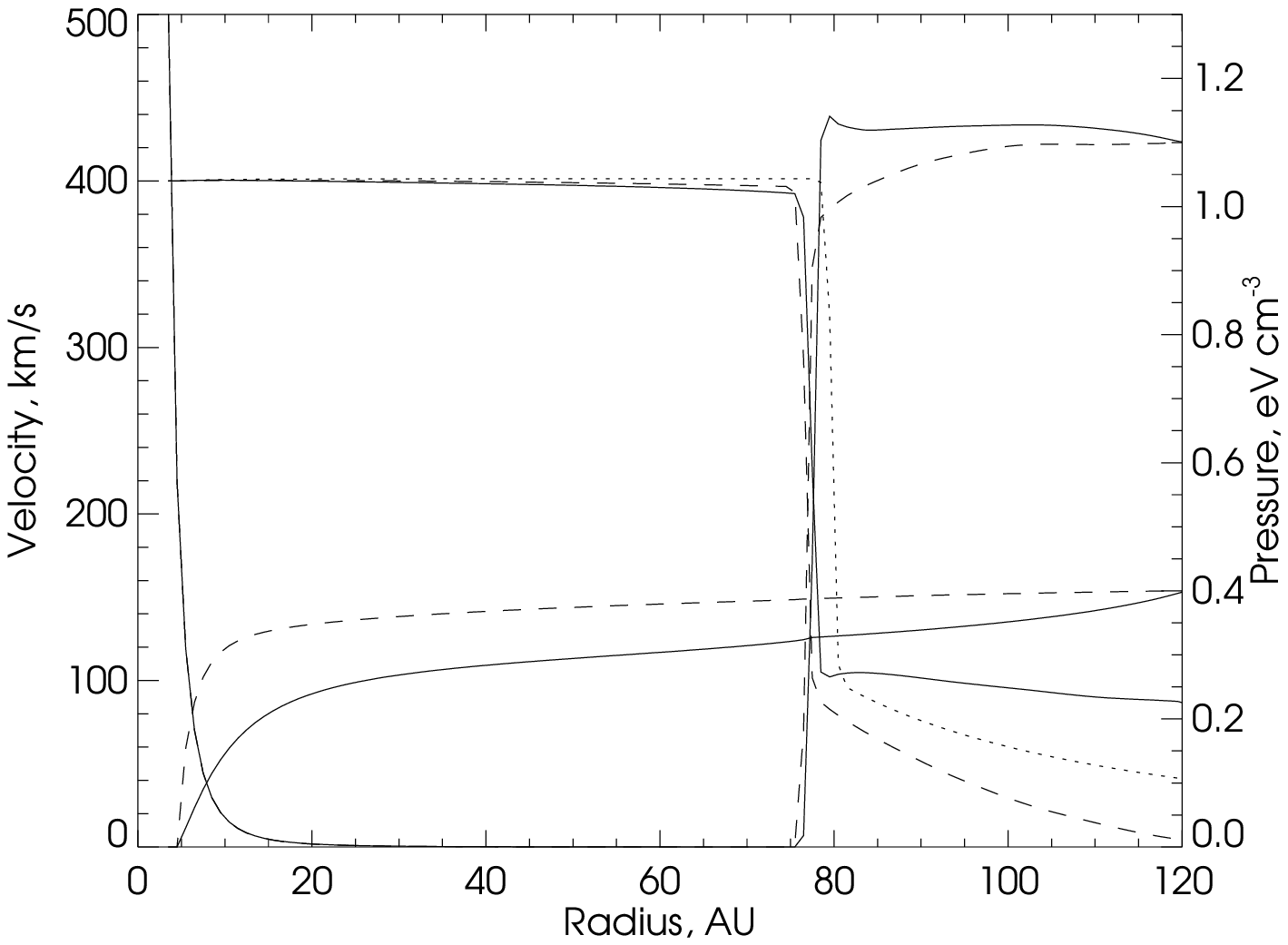}
\figcaption{Radial wind speed and pressure, and GCR pressure at $0^0$ latitude
(solid lines) and at $90^0$ (dashed lines) for $A>0$.
Lines showing the wind speed begin on the left at 400 km/s, $P_g$ lines reach
the outer boundary value of 1.1 $\mbox{eV cm}^{-3}$ and the $P_c$ lines reach 
0.4 $\mbox{eV cm}^{-3}$ at the external boundary.
The dotted line is the latitude-independent wind speed without the cosmic rays.}
\end{figure}
Figure 4 shows equatorial and polar thermal and GCR pressures and the radial
wind velocities for the present epoch ($A>0$).
The termination shock has moved inward between 3 and 4 AU.
A wide shallow cosmic-ray shock precursor is visible in the supersonic wind.
Notice that the wind is faster at the equator as a result of compression and
slower near the pole due to expansion.
Equatorial gas pressure build up is due to both compression and conservation of
momentum downstream of the shock (the sum, $P_g+P_c$, tends to remain nearly a
constant).
The pressure peak just downstream of the shock at the equator corresponds to a
region of large $u_{\theta}$ at slightly higher latitude and seem to balance the
dynamic pressure of the equatorward flow.
However, it could possibly be an artifact of the solution.
The largest dynamic pressure, $\rho u_{\theta}^2/2$, associated with the
latitudinal flow is about 0.013 $\mbox{eV cm}^{-3}$ while the pressure peak is
0.08 $\mbox{eV cm}^{-3}$ above the average gas pressure in this region.
For $A<0$, wind velocity and thermal pressure profiles at the pole look similar
to those at the equator for $A>0$ and the peak is present, too.

\section{Discussion}
With a series of tests we have been able to verify that the $\theta$ gradients
of the cosmic-ray pressure are responsible for the solar wind flow configuration
we observed.
We were able to produce a general flow structure similar to that shown in Figure
1 by artificially placing a constant in time cosmic-ray pressure gradient that
varied in latitude only.
In turn, large lateral gradients are due to difference in radial drift velocity
between the pole and the equator, in agreement with Jokipii \& Davila (1981).
Because latitudinal drifts and diffusion tend to smooth out these gradients the
wind will be less affected.
The fact that the shock is still spherical remains unexplained, it could be
due to our pressure boundary conditions that are spherically symmetric.
\par
We have found that the system continues to evolve on time scales comparable to
the 11-year solar cycle.
We expect the heliospheric configuration to oscillate between states similar to
those shown on Figures 1a-1b and 2a-2b. 
We have not seen the system reach a steady state under stationary boundary
conditions when the cosmic-ray pressure is comparable to the thermal pressure at
the external boundary.
Notice a large positive radial gradient in $P_g$ in the polar region in the
$A>0$ case (Figure 4).
There the subsonic wind is propagating away from the Sun against this gradient
and can be quite easily turned backward.
Such an inflow would require specifying two extra conditions at the outer
boundary.
\par
Should an inflow occur, the magnetic field is dragged along the lines
separating the external flow and the solar wind.
Because the field offers no resistance, both $B_{\theta}$ and $B^m_{\theta}$ may
become quite large.
To see if this effect has physical significance we plan to add magnetic force to
the solar wind equations.
In addition, anomalous cosmic rays should contribute significantly to the
dynamics of the system; we plan to include them in a later publication.

\acknowledgments
This work was supported, in part, by NASA under grants NAG5-6620, NAG5-7793 and
by the NSF under grant ATM9616547.

{}

\end{document}